# *FINDING AN EFFECTIVE CLASSIFICATION TECHNIQUE TO DEVELOP A SOFTWARE TEAM COMPOSITION MODEL*

Abdul Rehman Gilal [1], Jafreezal Jaafar [2], Luiz Fernando Capretz [3], Mazni Omar [4], Shuib Basri [5], Izzatdin Abdul Aziz [6]

[1, 2, 5, 6] *Department of Computer and Information Sciences, Universiti Teknologi PETRONAS, Malaysia*
[1] *Department of Computer Science, Sukkur Institute of Business Administration, Pakistan*
[3] *Western University, London, Canada*
[4] *School of Computing, Universiti Utara Malaysia*

[1]a-rehman@iba-suk.edu.pk, [2]jafreez@utp.com.my, [3]lcapretz@uwo.ca, [4]mazni@uum.edu.my,
[5]shuib_basri@utp.edu.my, [6]Izzatdin@utp.edu.my

*Abstract*— Ineffective software team composition has become recognized as a prominent aspect of software project failures. Reports from results extracted from different theoretical personality models have produced contradicting fits, validity challenges, and missing guidance during software development personnel selection. It is also believed that the technique/s used while developing a model can impact the overall results. Thus, this study aims to: 1) discover an effective classification technique to solve the problem, and 2) develop a model for composition of the software development team. The model developed was composed of three predictors: team role, personality types, and gender variables; it also contained one outcome: team performance variable. The techniques used for model development were logistic regression, decision tree, and Rough Sets Theory (RST). Higher prediction accuracy and reduced pattern complexity were the two parameters for selecting the effective technique. Based on the results, the Johnson Algorithm (JA) of RST appeared to be an effective technique for a team composition model. The study has proposed a set of 24 decision rules for finding effective team members. These rules involve gender classification to highlight the appropriate personality profile for software developers. In the end, this study concludes that selecting an appropriate classification technique is one of the most important factors in developing effective models.

## I. INTRODUCTION

The software development industry has been involved in detrimental situations where only 6% of software under development is being delivered on time and on budget [1]. This community has been underestimated because they were thought to be less productive in creating software that can live up to its original expectations, due to the cost and frustration of failed or underperforming software. Some people think that they can only credit luck when software development projects and their performance succeed. To change this myth, several studies were conducted to discover the detrimental factors in software development [2]–[5]. Based on the identified factors, ineffective team composition appeared to be one of the important aspects of failure [6], [7]. In this study the term 'team' refers to a number of people who are correspondingly skillful and strive together to meet a common purpose. The criterion of team composition in software development projects has been mainly based on the technical skills of team members. However, a team can function most ideally if the technical (hard) skills are combined with non-technical (soft: social or personality) skills [8]. In the same vein, Dingsøyr and Dybå [9] maintained that isolation of either skill (technical or social) can be one of the reasons for poor software development. It is also believed that the consideration of technical skills of developers can be advantageous as long as software developers are also evaluated in terms of their personality traits, a soft skill, to determine whether they can work cooperatively with other team members [10]. Personality refers to an internal psychological pattern, such as feelings and thoughts, that shape the behavior of a person [11]. Including personality-based skills can create a healthy behavior among employees, which can lead to overall project success. If this is improperly managed it can also cause damage within project development.

A plethora of research has been carried out in the past to explore the key importance of team composition and personality types in software development [11]–[14]. However, the personality types that are useful and beneficial for ideal and effective teamwork are still not well-defined for practitioners and researchers [15]–[18]. Further, the results, extracted from different theoretical personality models, have produced contradictory fits, validity challenges, and missing guidance for software development personnel selection. For instance, according to da Silva et al. [37], different results were obtained when models were used in practice to compose teams. Moreover, other researchers believe that the numerous models that have been suggested for team composition have been shown not to yield positive results by scholars and organizations [9], [11], [19]–[21]. This raises another challenge for researchers to identify and solve: the reasons for contradictory fits. For example, the problem may exist in team composition models where different researchers accentuate different personality types for effective teamwork, without adequately considering their suitable and effective role in actual teamwork [22]. The contradictory results may also be due to different experimental setups [23]. Importantly, it may also be that the selection of classification techniques or the technique used for developing a model may have impacted the results [24]. Therefore, choosing and adopting suitable techniques is essential to effectively meet the goals [25].

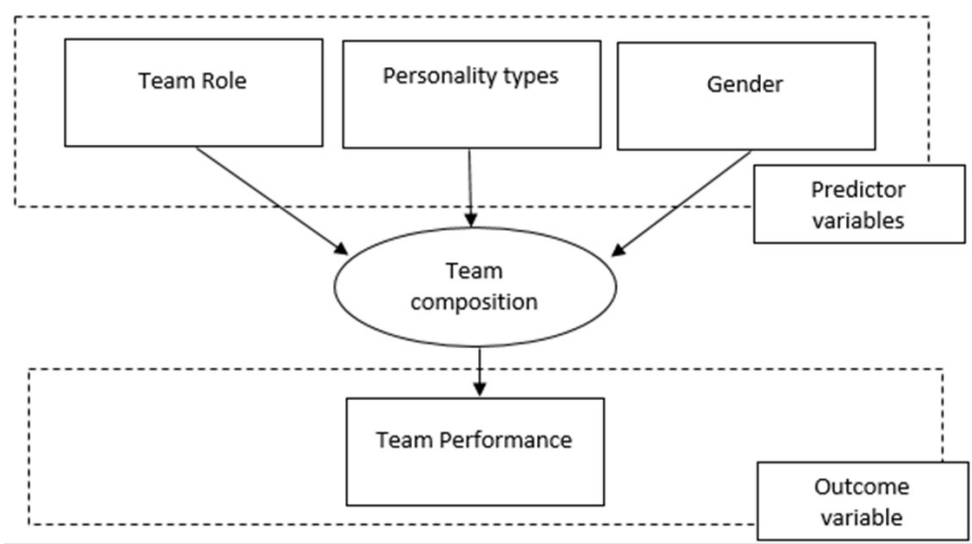

Fig. 1. Study Framework

Based on this situation, this study aimed to 1) discover an effective classification technique and 2) develop a model for software development team composition. The model was developed by considering software team roles, personality types, and gender as predictor variables for team effectiveness. Figure 1 above illustrates the study framework that shows the relationship between predictors and outcome variables. Moreover, while the combination of team roles and personality types has been explored by a few authors [26]–[29], gender has been ignored [1]. Gender is considered in this study because the increasing complexity of software development demands different personality profiles [30] and the personality profiles of male and female developers can never be same [31]. The field of psychology has had a great number of discussions about differences in personality possessed by males and females [32]. Males and females are categorized on different scales of traits in psychology, but personality-based studies in Software Engineering (SE) have dealt with them as identical. This is why past studies have suggested that the gender factor should be focused on in personality based research in software development [33]. According to Jayne and Heather [34], issues emerge when personality is interpreted without considering gender differences. It is also believed that not considering gender is one of the strong limitations in personality research.

The major contribution of this study is to offer a model for software development team composition. The implication of the developed model is to provide a mechanism for human resource managers, decision makers or leaders to compile an efficient team equipped with hard and soft skills. The results predicated from the model may also be useful to managers as preventive guidelines which can help them to avoid assigning the wrong type of individual to a team. Additionally, this study includes a gender based personality composition which also highlights the relationships between team roles. This approach will produce a better level of understanding of the fit between personality and software development roles. In this way, human capital can potentially be correlated to specific jobs.

The following section presents a related work on software development teams, data mining techniques and software development and personality to establish the basis for study variables. Once the groundwork has been established, the methodology section discusses the data collection experiments with detailed procedures of model design and development. The results and discussion section appears after the methodology section to highlight the important findings of the model development. The model validation section is especially kept in order to highlight the evolution of the model. Finally, threats to validity section focuses on the limitations of the model.

## II.     RELATED WORK

McConnell [35], Linberg [3], Demarco and Lister [36], and Nelson [37] have stated that information technology (IT) professionals and software development teams are not being respected. Even upper management and users question the unnecessarily high risk of failure for software development projects. McConnell et al. (1996) also added that many organizations reported that the number of software projects that fail to be implemented, created, or used to their fullest potential is far greater than the software projects that improve organizational performance. These days software demand is increasing rapidly in every field. At the same time, the development of successful software is decreasing. Various factors have been identified that affect the development process of software success as a whole. Among these factors, the formation of an unsuitable team was found to have the most impact [38], [39]. Based on research by Bell [40], team composition is a formation of team members according to their contributions in influencing the team processes and outcomes. Bell has also affirmed that the team composition research that has been carried out in past studies can be categorized into three types: (1) team member characteristics (e.g., member abilities, demographics, personality traits, and number of team members), (2) measurement of team member characteristics, (3) and team composition developed using an analytical perspective. As a matter of fact, team composition will gradually also be based on the technicality of the work. Nevertheless, both social norms and technicality have now been integrated by software development. Based on Capretz and Ahmed [14], hard (technical) skills and soft (non-technical: personality) skills need to be combined in order to create an ideal team. Dingsøyr and Dybå [9] also supported this idea when they stated that the reason for poor software development is the isolation of one of the skills – technical or social. In addition, technical skills of developers are believed to be able to give advantages as long as the soft skills, or personalities, of software developers are also being examined in order to ensure that they can work cooperatively with other team members [13].

In the social sciences, many research studies have explored personality and gender, either collectively or separately, to address the grave problems of teamwork in organizations, and they have achieved great success. However, this problem is still persistent in the field of software development since few researchers have ever tried to test personality and gender in assessing the suitability of the team available for software development. In this regard, Richards and Busch [41], Gilal et al., [27], and Rehman et al. [42] also asserted that maturity level has yet to considered in software development research. In the same vein, Trauth [33] also recommends that the theoretical work on software development needs improvement.

**2.1 Common Data Mining Techniques for Classification**

Data mining refers to the procedure of analyzing and discovering new and potentially interesting arrays of available data [43]. Data mining is also considered as a key source to unearth previously undiscovered new knowledge. The goals of data mining can be put into two broad categories: prediction and description. Prediction is used to find patterns that are put into practice to foresee unseen future growth and outcomes. Description signifies discovered patterns that can be understood and conceived by the users. As data mining is carried out using complex sets, it involves myriad models to obtain the desired goals and tasks. This view can be gained from the following figure that not only shows data mining models but also associated tasks.

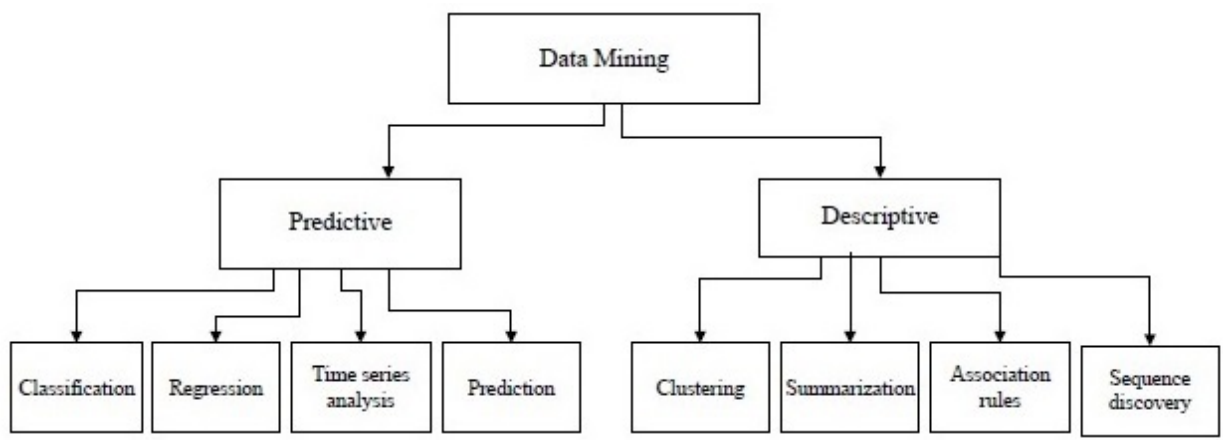

Fig. 2. Data mining models and task [44]

Based on Fig 2, the models of data mining can be described as two types: predictive and descriptive. Historical data is used as a primary source to make predications for future outcome under the umbrella of the predictive data mining model, whereas descriptive data mining is employed so as to explore and identify the patterns and relationships in the data. To put this into different wording, the predicative data mining model is labeled as supervised learning, because the outcome of the data is assumed or foreseen prior to the data analysis. By contrast, the descriptive data mining model is labeled as unsupervised learning where the outcome is neither known nor predicted. Hence, the predictive data mining model is the prime focus of this study to develop understanding and to discover the relationship between input and target output based on historical data. Additionally, the differences between the observed and actual output can be minimized to an extent with the help of developed algorithms. Myriad kinds of data mining techniques have been followed until recently, and most of them are followed using classical statistics. Few of the data mining techniques are performed following artificial intelligence. The data mining techniques that are mainly put into use are decision tree [45], logistic regression [46], artificial neutral network (ANN) [47], rough set [48], and support vector machine (SVM) [49]. Each of these techniques has its own strengths and challenges that can be taken into account before implementing them in data analysis. The following Table I presents an overview of the strengths and challenges of data mining techniques.

<<TABLE I>>

The key advantage of statistical logistic regression lies in the fact that it is quite effective while handling multiple predictors of mixed data types that yield reliable binary outcomes. Despite this advantage, this technique is not free of problems. First, it requires extensive data to perform modeling, and secondly, a researcher should be well qualified in statistical and domain knowledge to operate this technique competently. ANN is another popular technique in the field of data mining that not only enables researchers to predict tasks but also to help in securing higher and more accurate performance. However, its darker side cannot be neglected, because performing this technique can cause the over-fitting error in modelling. Moreover, it is considered as a black-box technique that is less effective in assessing categorical data.

SVM data mining technique is also popular among researchers because of a quality that enables researchers to obtain satisfactory performance and accuracy based on normal distributed data. However, unlike ANN, this technique is not suitable when employed in analyzing categorical data. To cope with this problem in handling the categorical data, researchers mostly rely on a decision tree technique that is effective for analyzing mixed data in general and analyzing categorical data in particular. But this technique must be carefully considered before it is employed because of its drawbacks for generating a complex tree structure that cannot be fitted easily into the data. Rough set is a quite new technique as compared to all other techniques. It facilitates researchers in generating the IF-THEN rule that can be easily interpreted. It is also suited for both small sample sizes and categorical data types. But this technique is often criticized for generating excessive rules that cannot be followed easily while making a pattern interpretation. Additionally, this technique works on continuous data discretization that mostly reduces data knowledge representation.

Keeping in view the merits and challenges of data mining techniques, Kotsiantis [50] maintains that since data mining is by its nature an exploratory process, no data mining technique or learning algorithm can be considered the best suited technique to analyze different data sets and domains. In the same vein, Dreiseitl and Ohno-Machado [51] state that researchers need to know the nature of data and employ the techniques accordingly. However, a past literature review reveals the fact that three kinds of data mining techniques cum rough set, decision tree, and logistic regression can be employed to devise prediction models. These techniques were declared safe techniques for devising prediction models based on the data normality assumption, sample size, and type of data.

## 2.2 Software Development and Personality

Cruz et al. [11] did an extensive systematic literature survey covering forty years, and claimed that research regarding software engineering mostly uses the Myers-Brigg Type Indicator (MBTI) instrument to assess personality types . MBTI is used in 36 studies out of 75, which is 48%, while seven out of 75 studies (7%) used the Kersey Temperament Sorter (KTS). These two instruments (MBTI and KTS), which are built from Carl Jung's Personality Types Theory, were used in 43 out of 75 studies (57%). Apart from that, 14 out of 75 studies (19%) used tests like NEO-PI test that was constructed from the Five Factor Model (FFM) as well as the Big Five (BF) theory. Three out of 75 studies (4%) were not using any tests regarding personality, while the rest of the studies used diverse kinds of tests. According to Furham [52], both MBTI and BF personality tests are worthwhile when a researcher aims to examine behavioural and cognitive sides of individuals by correlating both of the scales. However, there are many proponents of MBTI in the domain of software engineering, as this theory has been widely used in past research studies [1], [27], [29], [30], [53]–[56]. Thus, keeping in view the wide acceptance of MBTI in terms of its effectiveness, the current study has used this theory.

MBTI primarily focuses on four pairs of the personality which can be further classified into 16 types. The four pairs are: Extroversion-Introversion (I-E), Sensing-Intuitive (S-N), Thinking-Feeling (T-F), and Judging-Perceiving (J-P). These four dimensions also provide sixteen possible combinations of personality types as shown in Table II.

<<TABLE II>>

Based on the performance and the score obtained, a person can be considered to be one of the 16 personality types cited in Table II. For instance, a person scoring higher on Introversion (I) than Extroversion (E); Sensing (S) than Intuition (N); Thinking (T) than Feeling (F); and Judging (J) than Perceiving (P) would be categorized as an ISTJ.

## III. METHODOLOGY

This study has used the data collected from Universiti Teknologi Petronas (UTP) for model development and validation. Academic setup was selected for data collection experiments because it seemed more feasible than industrial. There has always been a trade-off between controlled and realism in selecting the study experiment locations. Sjøberg et al., [57] state that conducting study experiments in the office environment may increase the realism but there are several validity threats (i.e., internal and conclusion validity threats) which may lead the study away from the goals. Similarly, Hornbaek [58] also extends that "in itself having students participate in an experiment may not matter to a study" (p. 27). Therefore, monitoring and controlling students under the laboratory or classroom setup is much easier than industrial settings and can provide true causes and effects [59].

The data was collected to find the impact of team role, personality types, and gender on team performance. UTP undergraduate students in the Computer and Information Sciences (CIS) department enrolled in SE course (i.e., offered in the Jan-2015 session), were involved in the study to be part of the experiment. They were given client-based projects to develop in teams. Teams were encouraged to work on a small-scale on real problems happening at UTP, for example: UTP Security Department Summon System, UTP co-curriculum online registration or UTP E-Clinic. The teams were comprised of only five team members with one team lead and four programmers. Actually, teams with fewer than six members were more effective because they were easier to manage and monitor [57]. Software development involves several roles during the development process: team leader, analyst, designer, programmer, and tester. However, UTP data collection only included team leader and programmer roles, as these roles are major and included in all type of software development methodologies [60]. These roles were chosen for the study due to the scope and limitation of time and resources. Moreover, students were given freedom to compose teams based on their own choices. The researchers also believe that allowing participants to choose their roles reduces conflict and increases comfort levels among the participants [61]. Teams were given only 12 weeks to submit the projects. The submitted projects were evaluated by the clients (i.e., a focal person from the particular department which is facing the issue) and external lecturers to measure the quality of software based on the requirements. Focal person evaluators were meant to check whether submitted projects satisfied their needs or not. On the other hand, external lecturers were supposed to evaluate the efficient development of the projects. It was believed that the quality of the developed project represented the performance of the team and the performance of the team represented the performance of individual or team members. For example, an individual will not be considered effective if the project received a failing grade. Based on the evaluators' results, only those teams were considered "effective" which obtained 80% or above on marks in project results; otherwise teams were called "ineffective." Moreover, the UTP dataset consisted of 105 participants with 50 male and 55 female participants. Table III summarizes the transformation of the variables in the study techniques.

<<TABLE III>>

Logistic regression, decision tree, and rough sets were applied under the Knowledge Discovery in databases (KDD) process to extract the results. In order to select the best technique for software development team composition, the study experiments were divided into three levels: pattern extractions, pattern performance evaluation, and selection. The patterns were extracted based on the appropriate techniques formula of technique, which are discussed below. Once the patterns were defined, they were then evaluated using a k(10)-fold method because the data size was small [62]. K-fold cross validation method is more precise than hold-out but it is computationally expensive. This is because, in k-fold cross validation method, the data set is divided into k subsets in which each subset is trained k-1 time and tested k times. Furthermore, the selection of technique was based on two parameters – pattern complexity and prediction accuracy – from the 10-fold method. In other words, the technique was selected if the patterns were less complex and prediction accuracy was higher than with other techniques. It should be noted here that, according to Bakar [63] and Hvidsten [64], 70% prediction accuracy is an acceptable accuracy for model development. Therefore, this study set 70% as the benchmark for an effective accuracy. At the end, the model was validated using an F1-score measure using following formula with 50% or above benchmark [65], [66]:

$$f1\_score = 2 * \frac{Precision * recall}{Precision + recall} \quad \text{----------------} \quad (1)$$

### A. Logistic Regression

This study outcome variable was dichotomous where linear regression could not ever fulfil the linearity assumptions. Thus, logistic regression was used because it is a non-linear function. Additionally, another reason to use logistic regression was the "data type of variables" used in this study. For instance, predictor variables have both types of data (continuous and categorical) and the outcome variable is categorical with only two classes (effective teams and ineffective teams). Lastly, logistic regression is free from data normality assumption. The following equation expresses the logistic regression:

$$Ln\left[\frac{P}{1-P}\right] = A + B_1X_1 + B_2X_2 + B_3X_3 \ldots \ldots B_nX_n \text{--}(2)$$

where A is a constant and $B_1$, $B_2$, …, $B_n$ are regression coefficients of the predictor $X_1$, $X_2$, …, $X_n$ variables, respectively, and can be interpreted as the estimated size of contribution of the corresponding predictor variable $X_i$ to the changes in outcome variable, p. Moreover, SPSS software was used to employ the logistic regression analysis on the data.

### B. Decision Tree

There are several decision tree learning algorithms in data mining: CLS [67], ID3 [68], C4.5 [45], and CART [69]. But C4.5 is the most popular algorithm because it has resolved the over-fitting problem by adding the pruning facility [70], which is the key drawback of ID3 and CART. The pruning process facilitates removing the least significant attribute to be classified for prediction. Usually, over-fitting occurs from noisy data which causes an exaggeration in prediction accuracy. Thus, this study used the C4.5 algorithm to employ the decision tree algorithm. The Waikato Environment for Knowledge Analysis (WEKA) tool was used for implementing the C4.5 algorithm. Basically, WEKA is a software tool which has a collection of several supervised and unsupervised data mining algorithms. J48 is a java-based implementation of the C4.5 algorithm in WEKA. The following steps were taken to run C4.5 in WEKA:

1) Select data set
2) Select decision tree algorithm (this study used J48 implementation of C4.5)
3) Evaluate the tree (cross-validation with k-fold)

### C. Rough Set Theory (RST)

RST uses lower and upper approximations of the original set, where the lower approximation denotes that domain objects definitely belong to the subset of interest. Whereas, upper approximation objects possibly belong to the subset of interest. The concept of RST is based on the concept of discernibility relationships of a decision table that comprises decision attributes (D) and conditions (C). The decision table represents the columns as the attributes and rows of the object or record of data. The information system (S), represented by the decision table, can be explained as S = {U, C, D}. In RST, the decision table acts as learning examples to enable for generation of decision rules [71] which can be in the following form:

$$IF\ (a1, v1)\ and\ (a2, v2)\ and…and\ (an, vn)\ THEN\ Class_j$$

where $a_i$, is the i-th attribute, $v_i$ its value and $Class_j$, is j-th decision class

The Rough Set Toolkit for Analysis of Data (ROSETTA) [72] tool was used to analyze the data for RST results. The tool was designed within the RST discernibility framework and it was integrated with the collection of RST algorithms. Thus, it is an efficient tool for analyzing research data based on the rough set approach. ROSETTA supports several basic algorithms for performing related actions of rough sets. For instance, data discretization (i.e., Boolean reasoning algorithm, manual discretization, and Naïve Scaler), reduction (i.e., Genetic algorithm, Johnson algorithm, and Holte's 1R), and classification (i.e., Batch classifier, standard voter, and object tracking voter). Therefore, ROSETTA generates results based on the following steps:

1) Data discretization
2) Reducts generation
3) Decision rules generation
4) Decision rules evaluation

Moreover, two heuristic search algorithms, Genetic Algorithm (GA) [73] and Johnson Algorithm (JA) [74], were used to find reducts and rules that can produce decision rules. Fewer reducts represent less complexity in the model. Hvidsten [64] states that GA is an effective method to search optimal solutions and to solve searching problems. In addition, Johnson [74] stated that JA invokes a variation of a simple greedy algorithm to compute a single reduct only

## IV. RESULTS AND DISCUSSION

Pearson [75] states that prediction accuracy is one of the most important factors for measuring the performance of a model. The selection of the technique was based on the prediction accuracy benchmark, which was 70%. Results showed that logistic regression was not a suitable option for model development because the prediction obtained accuracy was 67.6%. Similarly, the decision tree algorithm produced reasonable prediction accuracy results with 70.48%. This technique could not be selected because the rough set technique earned higher prediction accuracy than did the decision tree. For example, JA algorithm in the RST technique produced results with 79.04% prediction accuracy. In the same vein, GA algorithm in RST technique earned 75.23% prediction accuracy. Table IV summarizes the results of all techniques used in the study.

<<TABLE IV>>

Moreover, it was mentioned earlier in that among these two techniques, a technique with less complexity and higher prediction accuracy was to be selected. In this case, the JA algorithm appeared most suitable for this study because it produced 24 decision rules with 79.04% accuracy. Whereas, GA produced 48 decision rules with 75.23% prediction accuracy. Therefore, this study picked up the JA decision rules to use for further discussion on the model development.

The JA algorithm of RST was selected after validating and comparing it with other techniques. This algorithm produced 24 decision rules in the first experiment. Decision rules were tested for prediction accuracy to use in future studies. These 24 decision rules contained 10 decision rules for the team leader role, 10 decision rules for programmer role and 4 general decision rules that could be applied to both team roles. Table V presents the decision rules with basic information on the left hand side (LHS) and on the right hand side (RHS) supports.

<<TABLE V>>

Prior to the discussion of rules, it seemed appropriate to explain the key terms mentioned in the table above. For example, the "Decision rule" can also be used as a statement of decision based on an "IF-THEN" state driven from the dataset. Each decision rule had two clauses: 1) Condition (LHS) and 2) Decision (RHS). In the table above, the statement before "=>" is called the IF-PART and the statement after this (i.e., "=>") is called the THEN-PART. The term "LHS support" refers to how many objects from the dataset match the if-statement. Whereas, the term "RHS support" shows how many objects match the "then-statement/then-part" from the dataset based on the if-statement. Moreover, RHS support will show the same number of objects if the decision part had only one decision to make. For example, rule number 1 (from Table V) had 1 in LHS support as well as RHS support. In this study, these types of rules are called uni-dimensional rules. However, RHS support divides the value into two numbers if the decision part had two decisions. For instance, decision rule no. 12 (i.e., "programmer AND introvert AND sensing AND judging => ineffective OR effective" from Table V) had "9, 5" RHS support. It mentions that the If-part of the rule had been influenced nine times in the ineffective class and five times in the effective class. These types of decision rules are called bi-dimensional because they produce two different dimensions for decisions.

Based on the results, fourteen decision rules can be used to find the independent, team leader role. Ten decision rules were dedicated to the team leader role and four could be used for both roles. In these fourteen rules, four rules (rules no 1, 8, 10, and 21) highlight that a team leader can produce effective outcomes if composed accordingly. According to these rules, a female leader produced effective outcomes if the team was composed of members with the sensing (S) personality trait (see rule no 1). This pair of personality (S-N pair) is used for information collection in which the sensing trait collects the information using the five senses. It means that female developers with the sensing trait appeared to be suitable for the team leader role. Similarly, extrovert team leaders tend to have effective outcomes if they were composed with thinking + judging and intuiting + judging (see rules no 8 and 10). In other words, E + T + J and E + N + J had a high possibility to earn effective outcomes in software development projects. Lastly, male developers were found to have effective outcomes if they were composed with intuiting + perceiving personality traits. Furthermore, Gorla and Lam [22] found the E + N + T personality traits combination suitable for the team leader role. However, in this study, E + N or E + T were only suitable when these are composed with the J trait. In the same vein, this study also found that the sensing personality trait is suitable for female leaders; this was not found in the study of Gorla and Lam. On the other hand, Table V contained 11 decision rules that supported several different combinations which are ineffective in the team leader role. For example, previously in the descriptive analysis discussion, the finding showed that leaders with the introverted trait could not produce effective outcomes. But these decision rules showed that introverted trait leaders were not suitable if they were paired with sensing, thinking, and feeling personality traits (see rules no 4, 5, and 6). Basically, decision rules no 2, 3, 4, 5, 6, 7, 9, 22, 23, and 24 in Table V were to have ineffective outcomes for the team leader role. In addition, the bi-dimension rule of the team leader role, decision rule no 9, is discussed later in this section during discussion of the bi-dimension rules. Bi-dimension rules were summarized to be either effective or ineffective class based on their LHS and RHS coverage in the dataset.

In the same way, fourteen decision rules were formed to make a decision in selecting for the programmer role. Based on the basic dimension categories, eight decision rules were uni-dimensional and six decision rules were bi-dimensional. In eight uni-dimensional rules for programmer, rules no 13, 18, 20, and 21 in Table V supported effective outcomes for the programmer role. For example, a male programmer was found to be effective if he had introvert + perceiving personality traits. Similarly, extrovert personality programmers also appeared effective if their profile combined intuiting + perceiving and feeling + perceiving personality traits. One thing was discovered in this stage: the perceiving personality trait is suitable for the programmer role. Moreover, on the other hand, decision rules no. 16, 22, 23, and 24 in Table V supported ineffective outcomes for the programmer role. Decision rules no. 22, 23, and 24 were general rules for both roles: team leader and programmer. So far, only decision rule no 16 was dedicated to the programmer role and it indicates the personality type that can cause ineffective outcomes. In this decision rule, it is clearly highlighted that the INFJ personality type is not suitable for female programmer. Additionally, ISFP and ESFJ personality types were found not suitable for both roles: team-lead and programmer.

It is important to note that all of these bi-dimension rules were categorized based on their LHS and RHS coverage in the dataset. LHS Coverage refers to the overall appearance of the "If-part" in the dataset by dividing LHS support with the total objects of the dataset (i.e., 105). On the other hand, the RHS Coverage term is similar to LHS coverage but it covers the "then-part" of the rule by dividing RHS support with the total classified objects of the dataset (i.e., effective=45 and ineffective=60). Hence, based on the results, seven decision rules were bi-dimension, which divided the decision part into both effective and ineffective. Table V presents those bi-dimension rules with numbers 9, 11, 12, 14, 15, 17, and 19. Table VI keeps the same decision rule numbers to present the LHS and RHS coverage of bi-dimensional rules.

<<TABLE VI >>

Based on the LHS and RHS coverage, rules number 9 and 15 covered the same area in the dataset (i.e., LHS coverage=0.019048 and RHS=0.022222, 0.016667). In this case the effective class appeared higher than ineffective with 0.022222 coverage. Hence, both of these rules were dedicated to making the decision for the effective class. Whereas, all of the remaining rules: rules no. 11, 12, 13, 14, 17, and 18 were used for deciding ineffective class due to a higher appearance in the results.

## V.  MODEL EVALUATION

The F1-score is also called a balanced or weighted harmonic mean of precision and recall. In order to find precision, TP (True Positive) instances should be divided with positive predicted values (PPV): True Positive / (True Positive + False Positive). In this study, based on the obtained results, the precision measure returned 100%, which shows the proportion of actual effectiveness returned by the model. Similarly, recall was measured by using the formula: True Positive / (True Positive + False Negative). Based on these results, 51.11% was found on the recall measure (i.e., (23/45)*100=51.11%). By using the recall measure, 22 instances that were actually effective or positive were found to have been missed by the model. Hence, it was important to see the balanced harmonic mean of these measures to evaluate the performance. Table VII presents the results of model evaluation in a confusion matrix for computing F1-score.

<<TABLE VII>>

The F1-score was measured using equation 1, which was discussed in the methodology section: 2*((precision * recall)/ (precision + recall)). Based on the obtained results of precision and recall, the F1-score measure obtained 67.65% in return. Whereas, the study baseline for accepting the performance of the model was 0.5 or 50%, based on the F1-score. Therefore, in this study this model was considered acceptable because it obtained the benchmark level of prediction accuracy and F1-score measures.

## VI.  THREATS TO VALIDTY

First of all, the results of this study can be used for finding an effective personality composition for individuals in the roles of team leader and programmer. However, personality is a complex term which is vague in nature and can be affected by several internal and external factors. Thus, generalizing these results remains the main concern for the validity of the study. For instance, results from this study may not be generalized with other Malaysian universities without a cross validation. In order to generalize them, the results can be expanded by studying students in other universities in Malaysia and as well in other multicultural settings for more rules. Secondly, the study experiments were conducted within an academic setting, which limits the results for industrial settings; this validity threat was controlled by considering the age of participants. Due to the fact that researchers in the psychology domain claimed that personality is an inherited factor. Researchers agree that persons between the ages of 20-50 years have stable and consistent personality types [76]. Hence, the required minimum age of participants was 20 years in this study to control the validity threats. To validate the results further, data from industrial settings can also remove the threats of acceptance or generalization. Thirdly, team leader and programmer roles were adjusted to compose personality equations for effective team compositions. Usually, the persons in the roles of team leader and programmer also work with testers and designers. In this case, the results were restricted to be used within the teams which involve roles other than those of team leader and programmer. Thus, the results can be enriched if these are trends for the other roles: testers or designers. Furthermore, only MBTI-based personality types are offered in the model, which can be one of the limitations. Hence, other studies can include new rules based on personality assessments other than MBTI, such as the Big Five or Keirsey Temperament Sorter.

## VII. CONCLUSION

There are several important factors for contradicting results on software development models, and selecting appropriate techniques for model development is one of them. This study used three different techniques (logistic regression, decision tree, and RST) on the same data. Yet, each technique returned different results. In fact, logistic regression results could not meet the satisfaction level of model development. Hence, using only one technique to develop a model may lead to biased and unwanted outcomes. The results also suggest that the model validation process will produce effective results if checks are increased. For example, C4.5 and GA techniques were also found to be effective if they were only used for prediction accuracy checks. The conclusion of the first objective can be made by the statement that a model development technique should be selected carefully because it can cause different results during implementation. Furthermore, team members' personality plays an important role or effective team composition. Ignorance of personality variable during team role assignment may impact the overall results of software development. The study results also support a claim that male and female developers cannot be categorized under the same personality profile for the same role.

*Journal of Software: Evolution and Process*, 29(10), DOI: 10.1002/smr.1920, Wiley, October 2017.[27]  Gilal AR, Jaafa J, Omar M, Tunio MZ. Impact of personality and gender diversity on software development teams' performance. International Conference on Computer, Communications, and Control Technology (I4CT), 2014 (pp. 261-265). IEEE.

[28]  Capretz LF, Ahmed F, da Silva FQ. Soft sides of software. Information and Software Technology. 92,92-94, 2017. DOI: 10.1016/j.infsof.2017.07.011.

[29]  mar M, Syed-Abdullah SL. Identifying effective software engineering (SE) team personality types composition using rough set approach. International Symposium on Information Technology (ITSim), 2010 (Vol. 3, pp. 1499-1503). IEEE.

[30]  Varona D, Capretz LF, Piñero Y, Raza A. Evolution of software engineers' personality profile. ACM SIGSOFT Software Engineering Notes. 2012; 37(1):1-5. DOI: 10.1145/2088883.2088901.

[31]  Whipkey KL. Identifying predictors of programming skill. ACM SIGCSE Bulletin. 1984; 16(4):36-42.

[32]  Colins OF, Fanti KA, Salekin RT, Andershed H. Psychopathic personality in the general population: Differences and similarities across gender. Journal of personality disorders. 2017; 31(1):49-74.

[33]  Trauth EM. Theorizing gender and information technology research. In Human Computer Interaction: Concepts, Methodologies, Tools, and Applications 2009 (pp. 2309-2315). IGI Global.

[34]  Stake JE, Eisele H. Gender and personality. InHandbook of gender research in psychology 2010 (pp. 19-40). Springer New York

[35]  McConnell S. How to defend an unpopular schedule [software development projects]. IEEE Software. 1996; 13(3):118-120.

[36]  DeMarco T, Lister T. Peopleware: productive projects and teams. Addison-Wesley; 2013.

[37]  Nelson RR. IT project management: Infamous failures, classic mistakes, and best practices. MIS Quarterly executive. 2007; 6(2):67-78.

[38]  Zarzu C, Scarlat C, Falcioglu P. Team composition and team performance: achieving higher quality results in an international higher education environment. International Conference on Management, Knowledge and Learning, 2013 (pp. 1321-1328).

[39]  Mazni O, Syed-Abdullah SL, Hussein NM. Analyzing personality types to predict team performance. International Conference on Science and Social Research (CSSR), 2010 (pp. 624-628). IEEE.

[40]  Bell ST. Deep-level composition variables as predictors of team performance: a meta-analysis. Journal of applied psychology. 2007; 92(3):595-615.

[41]  Richards D, Busch P. Knowing-doing gaps in ICT: gender and culture. VINE. 2013; 43(3):264-295.

[42]  Rehman M, Mahmood AK, Salleh R, Amin A. Mapping job requirements of software engineers to Big Five Personality Traits. International Conference on Computer & Information Science (ICCIS), 2012 (Vol. 2, pp. 1115-1122). IEEE.

[43]  Fayyad U, Piatetsky-Shapiro G, Smyth P. The KDD process for extracting useful knowledge from volumes of data. Communications of the ACM. 1996; 39(11):27-34.

[44]  Dunham MH. Data mining: Introductory and advanced topics. 2003. Pearson Education.

[45]  Quinlan JR. C4. 5: Programming for machine learning. Morgan Kauffmann. 1993; pp. 38.

[46]  Cox DR. The regression analysis of binary sequences. J R Stat Soc Ser B. JSTOR; 1958;215–42.

[47]  Rosenblatt F. The perceptron: A probabilistic model for information storage and organization in the brain. Psychol Rev. American Psychological Association; 1958;65(6):386.

[48]  Pawlak Z. Rough sets. Int J Comput Inf Sci. Springer; 1982;11(5):341–56.

TABLE I. SUMMARY OF DATA MINING TECHNIQUES

| No | Data Mining Techniques | Author (s) | Advantages | Disadvantages |
|---|---|---|---|---|
| 1 | Logistic Regression | Ayer et al., [77]; Hosmer & Lemeshow, [78] | • No assumption of linearity<br>• Able to make prediction based on multivariate predictor variables of mixed data types and produce categorical binary outcome<br>• Easy to identify important predictor variables | • Requires sufficient data (i.e., at least 10 cases for 1 predictor)<br>• Requires domain and statistical knowledge |
| 2 | Artificial Neural Network (ANN) | Ayer et al., [77]; Dunham, [44]; Kotsiantis, [50] | • No assumption of linearity<br>• Able to obtain high performance accuracy | • Requires extensive data training<br>• Tends to result in over-fitting<br>• Black-box approach, thus difficult to interpret relationships amongst variables investigated<br>• Not suitable for categorical data |
| 3 | Support Vector Machines (SVM) | Kotsiantis, [50]; Olson & Delen, [79] | • No assumption of linearity<br>• Able to obtain high performance accuracy | • Requires much of the training data<br>• Not suitable for categorical data types |
| 4 | Decision Tree | Dunham, [44]; Kotsiantis, [50]; Mazni et al., [80] | • No assumption of linearity<br>• Easy to visualize<br>• Suitable for mixed data type (i.e., continuous, categorical) | • Can generate decision trees of various complexities<br>• Tends to result in over-fitting |
| 5 | Rough Set | Hui, [81]; Olson & Delen, [79] | • No assumption of linearity<br>• Generates IF-THEN rules, which are easier to interpret<br>• Suitable for categorical data type<br>• Can deal with small sample size | • Can generate excessive decision rules, thus it is hard to interpret<br>• Data must be discretized |

TABLE II. THE 16 MBTI PERSONALITY TYPES

| ISTJ (1) | ISFJ (2) | INFJ (3) | INTJ (4) |
| --- | --- | --- | --- |
| ISTP (5) | ISFP (6) | INFP (7) | INTP (8) |
| ESTP (9) | ESFP (10) | ENFP (11) | ENTP (12) |
| ESTJ (13) | ESFJ (14) | ENFJ (15) | ENTJ (16) |

TABLE III. CONTROL OVER STUDY VARIABLES

| Variable | Input |
|---|---|
| **Predictor** | |
| 1. **Member Role** | 1=team lead<br>2=programmer |
| 2. **IE** | 1=introvert<br>2=extrovert |
| 3. **SN** | 1=sensing<br>2=intuiting |
| 4. **TF** | 1=thinking<br>2=feeling |
| 5. **JP** | 1=judging<br>2=perceiving |
| 6. **Gender** | 1=Male<br>2=Female |
| **Outcome** | |
| 1. **Team performance** | 0= Ineffective<br>1= Effective |

TABLE IV. SUMMARY OF DATA MINING TECHNIQUES

| Criteria | Techniques | | | |
|---|---|---|---|---|
| | **Regression** | **Decision Tree** | **Rough Set** | |
| | Binominal regression | C4.5 | GA | JA |
| **Association Patterns** | -Show significant Predicator Variables<br>-acquire statistical knowledge to implement | -Visualize significant relationships in tree form<br>-easy to determine effectiveness and ineffectiveness | -IF-THEN rules<br>-easy to use, make finite machine, make relational algebra | -IF-THEN rules<br>-easy to use, make finite machine, make relational algebra |
| **Prediction Accuracy** | 67.6% | 70.48% | 75.23% | **79.04%** |
| **Decision Rules** | ----- | 8 rules in pruned tree | 48 decision rules | **24 decision rules** |
| **Decision (accepted / rejected )** | Rejected | Rejected | Rejected | **Accepted** |

TABLE V. JA ALGORITHM DECISION RULES

| S.no | Decision Rule | LHS Support | RHS Support |
|---|---|---|---|
| 1 | team leader AND sensing AND female => effective | 1 | 1 |
| 2 | team leader AND perceiving AND male => ineffective | 4 | 4 |
| 3 | team leader AND intuiting AND feeling AND perceiving => ineffective | 1 | 1 |
| 4 | team leader AND introvert AND thinking => ineffective | 7 | 7 |
| 5 | team leader AND introvert AND sensing => ineffective | 2 | 2 |
| 6 | team leader AND introvert AND feeling => ineffective | 1 | 1 |
| 7 | team leader AND feeling AND male => ineffective | 3 | 3 |
| 8 | team leader AND extrovert AND thinking AND judging => effective | 4 | 4 |
| 9 | team leader AND extrovert AND intuiting AND thinking AND perceiving => effective OR ineffective | 2 | 1, 1 |
| 10 | team leader AND extrovert AND intuiting AND judging => effective | 4 | 4 |
| 11 | programmer AND thinking AND judging => effective OR ineffective | 32 | 13, 19 |
| 12 | programmer AND introvert AND sensing AND judging => ineffective OR effective | 14 | 9, 5 |
| 13 | programmer AND introvert AND perceiving AND male => effective | 8 | 8 |
| 14 | programmer AND introvert AND judging AND male => effective OR ineffective | 14 | 5, 9 |
| 15 | programmer AND introvert AND intuiting AND perceiving AND female => effective OR ineffective | 2 | 1, 1 |
| 16 | programmer AND introvert AND intuiting AND feeling AND judging AND female => ineffective | 2 | 2 |
| 17 | programmer AND extrovert AND sensing AND thinking => effective OR ineffective | 14 | 6, 8 |
| 18 | programmer AND extrovert AND intuiting AND perceiving => effective | 6 | 6 |
| 19 | programmer AND extrovert AND intuiting AND judging => ineffective OR effective | 15 | 9, 6 |
| 20 | programmer AND extrovert AND feeling AND perceiving => effective | 5 | 5 |
| 21 | intuiting AND perceiving AND male => effective | 10 | 10 |
| 22 | introvert AND sensing AND perceiving AND female => ineffective | 2 | 2 |
| 23 | introvert AND sensing AND feeling AND perceiving => ineffective | 1 | 1 |
| 24 | extrovert AND sensing AND feeling AND judging => ineffective | 7 | 7 |

TABLE VI. LHS AND RHS COVERAGE OF BI-DIMENSION RULES

| Decision rule no | LHS coverage | RHS Coverage | Final Decision |
|---|---|---|---|
| 9 | 0.019048 | 0.022222, 0.016667 | Effective |
| 11 | 0.304762 | 0.288889, 0.316667 | Ineffective |
| 12 | 0.133333 | 0.15, 0.111111 | Ineffective |
| 14 | 0.133333 | 0.111111, 0.15 | Ineffective |
| 15 | 0.019048 | 0.022222, 0.016667 | Effective |
| 17 | 0.133333 | 0.133333, 0.133333 | Ineffective |
| 19 | 0.142857 | 0.15, 0.133333 | ineffective |

TABLE VII. CONFUSION MATRIX FOR PERFORMANCE EVALUATION

| | | Predicted | | |
|---|---|---|---|---|
| | | Ineffective | Effective | |
| **Actual** | Ineffective | 60(TN) | 0(FP) | 100 |
| | Effective | 22(FN) | 23(TP) | **51.11(Recall)** |
| | | **73.1707** | **100 (Precision)** | **79.0476(Accuracy** |